\documentclass[12pt]{article} 
\begin{document}
\title{\bf Dirty black holes: Spacetime geometry and near-horizon symmetries}
\author{A J M Medved, Damien Martin, and Matt Visser\\
School of Mathematical and Computing Sciences\\
Victoria University of Wellington\\ 
PO Box 600, Wellington\\ 
New Zealand}
\date{14 February 2004; 22 February 2004}
\maketitle

\begin{abstract}
  
  We consider the spacetime geometry of a static but otherwise generic
  black hole (that is, the horizon geometry and topology are not
  necessarily spherically symmetric).  It is demonstrated, by purely
  geometrical techniques, that the curvature tensors, and the Einstein
  tensor in particular, exhibit a very high degree of symmetry as the
  horizon is approached.  Consequently, the stress-energy tensor will
  be highly constrained near any static Killing horizon.  More
  specifically, it is shown that --- at the horizon --- the
  stress-energy tensor block-diagonalizes into ``transverse'' and
  ``parallel'' blocks, the transverse components of this tensor are
  proportional to the transverse metric, and these properties remain
  invariant under static conformal deformations.  Moreover, we
  speculate that this geometric symmetry underlies Carlip's notion of
  an asymptotic near-horizon conformal symmetry controlling the
  entropy of a black hole.

\bigskip

\centerline{gr-qc/0402069}

\bigskip

\centerline{{\sf damien.martin@mcs.vuw.ac.nz}, 
            {\sf joey.medved@mcs.vuw.ac.nz}}
\centerline{ {\sf matt.visser@mcs.vuw.ac.nz}}
\end{abstract}

\enlargethispage{250pt}
\clearpage

\section{Introduction}
\def\tr{\hbox{\rm tr}}
\def\implies{\Rightarrow}
\def\conv{\hbox{\rm conv}}
\def\Re{ {\cal R} }
\def\half{{{1\over2}}}
\def\d{{\mathrm{d}}}

Roughly 30 years has elapsed since the initial proposal that black
holes behave as thermodynamic systems; nonetheless, Bekenstein's black
hole entropy~\cite{BEK,HAW},
\begin{equation}
S_{BH} = {1\over 4}\left[{\rm Horizon\;area\;in\;Planck\;units}\right],
\end{equation}
is still one of the most puzzling (yet intriguing!)  concepts in
theoretical gravity~\cite{DAM}.  It will be a challenge to any
prospective fundamental theory to provide an unambiguous and universal
explanation for this entropy at the level of state counting
\cite{SMO}.  In spite of the relative success in this regard of, for
instance, string theory~\cite{STR-V} and loop quantum gravity
\cite{ASH}, it is safe to say that the microscopic origin of $S_{BH}$
remains a decidedly open question.

It seems likely that the microstates underlying black hole entropy
will only be fully understood at the level of quantum gravity.  On the
other hand, there is a growing suspicion that these states (whatever
they may be) are actually controlled by a {\em classically} inherited
symmetry~\cite{CAR3}.  This notion can, in large part, be attributed
to Strominger's realization~\cite{STR} that the black hole entropy in
three spacetime dimensions~\cite{BTZ} can be calculated by way of
Cardy's formula~\cite{CARDY}; notably, a formula which counts the
density of states for a two-dimensional conformal field theory.
Significant to this calculation was an earlier observation made by
Brown and Henneaux~\cite{BH}: the diffeomorphism invariance of
three-dimensional (anti-de Sitter) gravity can be manifested as a
two-dimensional conformal field theory that, in some sense, ``lives'' at
the boundary of the spacetime.

An obvious limitation of Strominger's work is that it directly applies
to only a three-dimensional theory of gravity. Nonetheless, progress
in four (as well as an arbitrary number of) dimensions of spacetime
has since been made by Carlip~\cite{CAR1,CAR2,CAR4} and Solodukhin
\cite{SOL}.  In these studies, the conformal field theory is now
regarded as  living at the black hole horizon.  Although the
horizon is not a true physical boundary (for instance, a free-falling
observer is not even aware of its existence), it is indeed a place
where boundary conditions can and should be set~\cite{CAR3}. Moreover,
when one considers issues of locality, the horizon is naturally
preferred over asymptotic infinity as the boundary that harbors the
relevant degrees of freedom.  Unfortunately, the precise choice of
conditions --- which, at this point, is somewhat ambiguous --- will
greatly influence any such determination of the entropy.  Hence,
substantial progress in this program will likely require a better
understanding of precisely what classical symmetry is at play. Most
recently, Carlip has suggested that the Einstein--Hilbert action
acquires a ``new'' asymptotic conformal symmetry in the neighborhood
of the horizon~\cite{CAR4}.  Although this seems intuitively correct
and is supported by the elegance of his calculation, this symmetry
still lacks the type of fundamental (classical) explanation that was
alluded to above.

Perhaps, the key to understanding Carlip's notion of a conformal
symmetry can be found in the Einstein field equations rather than the
action {\it per se}. To motivate this perspective, let us take note of
the following observation: any static (black hole) Killing horizon
with the property of spherical symmetry is known to possess the
following symmetry as the horizon is approached~\cite{DBH1,PAD}:
\begin{equation}
G^t{}_t \;-\; G^r{}_r \rightarrow 0 \;,
\label{X5}
\end{equation}
where $G_{\mu\nu}$ represents the Einstein (curvature) tensor, while
$t$ and $r$ are the usual Schwarzschild (or, in the case of ``dirty''
black holes, \footnote{In the current context, a dirty black hole is
  meant to imply a {\em generic} static and spherically symmetric
  spacetime for which a central black hole is surrounded by arbitrary
  matter fields~\cite{DBH1,DBH2,DBH3} . In the upcoming analysis, we
  specifically lift the condition of spherical symmetry.  For example,
  a ring of material might be placed around the equator of what would
  otherwise be a Schwarzschild black hole, distorting its horizon into
  an ovaloid shape.} Schwarzschild-like) temporal and radial
coordinates.  After imposing Einstein's field equations, one then
immediately obtains, at the horizon,
\begin{equation}
T^t{}_t \;-\; T^r{}_r = 0 \;,
\label{X6}
\end{equation}
where $T_{\mu\nu}$ is the stress-energy tensor.  Hence, restricting
attention to the $r$--$t$-plane (where all of the ``interesting''
near-horizon physics should presumably take place~\cite{CAR3}), we
have $T_\perp \propto g_\perp$.

The above observation is certainly interesting but, alas, the
condition of spherical symmetry seems quite restrictive.  [The
condition of staticity is also restrictive. However, for a slowly
evolving black hole, as long as the evolution rate is small compared
with the surface gravity, it could always be argued that the spacetime
is approximately static --- or, at the very least, approximately
stationary --- in the neighborhood of the horizon.]  The main purpose
of the current paper is to rectify this situation by establishing a
suitable analogue to equation (\ref{X5}) [and, hence, equation
(\ref{X6})] for a static but otherwise generic Killing horizon.

With the above discussion in mind, our aim is to develop a general
analysis for studying the geometric structure of a generic static
Killing horizon.  We begin, in Section 2, by using the natural time
coordinate in a static [(3+1)-dimensional] spacetime to slice this
geometry into space plus time.  Next, we twice employ the
Gauss--Codazzi and Gauss--Weingarten equations; allowing us to
decompose the spacetime Einstein tensor in terms of the geometrical
properties of a spatial $2$-surface embedded in a constant-time slice.
In Section 3, we then assume the existence of a Killing horizon and
consider the near-horizon limit of the generic formalism.  By way of
purely geometrical arguments, we are able to formulate a clear
description of the near-horizon geometry.  The Einstein tensor turns
out to indeed have a high degree of symmetry (at the horizon), and we
proceed to elaborate upon the implications of this outcome.  Extremal
horizons are discussed in Section 4, while the effect of conformal
transformations is discussed in Section 5.  Finally, Section 6
contains further discussion and a comment on future directions of
study.

We mention, in passing, that our analysis also enables an independent
and physically transparent verification of the ``zeroth law of black
hole thermodynamics'' (that is, the constancy of the surface gravity
\footnote{See, for instance, Wald~\cite{Wald}.} ) for static Killing
horizons. The main argument is presented in Section 3, with some
supporting calculations in an appendix.

\section{Generic static spacetimes}

\subsection{The (3+1)-geometry}

We will begin by considering  the geometry of a (3+1)-dimensional
spacetime which is constrained to be static but is otherwise
completely generic.  (Conditions as  appropriate for the existence of a
black hole Killing horizon will be imposed later on.) Note that the
methodology of this section is based largely on the techniques
of~\cite{Israel67,Israel68,Israel86,Hochberg-Visser,Hochberg-Visser-2}.

Given {\em any} static spacetime, one can always decompose the metric
into a block-diagonal form as follows~\cite{MTW,Hawking-Ellis,Wald}:
\begin{eqnarray}
\d s^2 &=& g_{\mu\nu} \; \d x^\mu \d x^\nu
\\
&=&
- N^2 \;\d t^2 + g_{ij} \; \d x^i \d x^j \;.
\end{eqnarray}
\underline{\underline{\em Notation:}} Greek indices run from 0--3 and
refer to the complete spacetime; whereas Latin indices in the middle
of the alphabet ($i$, $j$, $k$, \dots) run from 1--3 and refer to the
spacelike coordinates.  Also, Latin indices at the beginning of the
alphabet ($a$, $b$, $c$, \dots) will run from 1--2 and refer to
directions parallel to a soon-to-be-defined arbitrary spacelike
$2$-surface.  Furthermore (for an arbitrary geometrical object $X$),
we will use $X_{;\alpha}$ or $\nabla_{\alpha}X$
 to denote a spacetime covariant derivative,
$X_{|i}$ to denote a three-space covariant derivative, and $X_{:a}$ to
denote a two-space covariant derivative (which is always taken on the
aforementioned $2$-surface).  Finally, capitalized  Latin indices at the
beginning of the alphabet ($A$, $B$, $C$, \dots) will run from 0--1
and refer to the two directions perpendicular to the
arbitrary spacelike $2$-surface; essentially, the
$t$--$n$-plane defined by the time direction and the normal direction.

Let us next consider the three-geometry of space on a constant-time
slice. As it so happens, the property of staticity tightly constrains
the manner in which this three-geometry can be embedded into the
spacetime.  For instance, by way of a standard textbook~\cite[page
518]{MTW}, we have the following results:
\begin{eqnarray}
{}^{(3+1)}R_{ijkl} &=& 
{}^{(3)}R_{ijkl} \;, 
\\
{}^{(3+1)}R_{\hat tijk} &=& 0 \;,
\\
{}^{(3+1)}R_{\hat ti\hat tj}  &=&  
{N_{|ij}\over N}\;.
\end{eqnarray}
Here (and throughout), the ``hat'' on an index indicates that we are
looking at appropriately normalized components; for instance,
\begin{equation}
X_{\hat t} = X_t \; \sqrt{-g^{tt}} = {X_t\over\sqrt{-g_{tt}}} 
= {X_t \over N} \;.
\end{equation}
One can interpret this choice of normalization as using the
orthonormal basis attached to the fiducial observers (FIDOS).

Now decomposing the spacetime metric in terms of the spatial
$3$-metric, the set of vectors ${e^{\mu}{}_i}$ tangent to the time
slice [the triad or drei-bein], and the vector $V^\mu = {1\over N} \;
({\partial}/{\partial t })^{\mu}$ normal to the time slice, we have
\begin{equation}
{}^{(3+1)}g^{\mu \nu} = e^{\mu}{}_i \; e^{\nu}{}_j \; g^{ij} - V^\mu \; V^\nu
\;,
\end{equation}
which  can  be used to readily deduce the following contractions:
\begin{eqnarray}
{}^{(3+1)}R_{ij} &=& 
{}^{(3)}R_{ij} -  {N_{|ij}\over N} \;,
\\
{}^{(3+1)}R_{\hat t i} &=&  0 \;,
\\
{}^{(3+1)}R_{\hat t\hat t} &=&  
{ g^{ij} \; N_{|ij}\over N} 
 = {{}^{(3)}\Delta N\over N} \;,
\\
{}^{(3+1)}R &=&  {}^{(3)}R - 2 {{}^{(3)}\Delta N\over N} \;.
\end{eqnarray}
These results enable us to calculate the various components of the
Einstein tensor ({\it cf},~\cite[page 552]{MTW}):
\begin{eqnarray}
\label{E-static-stress-energy-b}
{}^{(3+1)}G_{ij} &=&  {}^{(3)}G_{ij}
 -  {N_{|ij}\over N} 
+ g_{ij} \; \left\{ {{}^{(3)}\Delta N\over N} \right\} \;,
\\
{}^{(3+1)}G_{\hat t i} &=& 0 \;,
\\
{}^{(3+1)} G_{\hat t\hat t} &=& + {1\over2} \;{}^{(3)} R \;.
\label{E-static-stress-energy-e}
\end{eqnarray}
It should be re-emphasized that this decomposition is generic to {\em
  any} static spacetime.

\subsection{The $3$-geometry}

Let us now (arbitrarily) choose a particular $2$-surface in the
constant-time slice and  utilize Gaussian normal coordinates in
the surrounding region; that is,
\begin{eqnarray}
 g_{ij} \; \d x^i \;\d x^j  =  \d n^2+  g_{ab} \; \d x^a \;\d x^b\;,
\end{eqnarray}
where $n={\hat n}$ represents the spatial direction normal to the
specified $2$-surface.  It can then be shown --- see for
example~\cite[page 514, equations (21.75-21.76) and page 516, equation
(21.82)]{MTW} ---  that
\begin{equation}
{}^{(3)}R_{abcd} = {}^{(2)}R_{abcd} - (K_{ac} K_{bd} - K_{ad} K_{bc} )
\;,
\label{X1}
\end{equation}
\begin{equation}
{}^{(3)}R_{\hat nabc} =  - (K_{ab:c}  - K_{ac:b}  ) \; ,
\end{equation}
\begin{equation}
{}^{(3)}R_{\hat na\hat nb} =  {\partial K_{ab} \over \partial n}  
+ (K^2)_{ab} \;,
\end{equation}
where the {\em extrinsic} curvature, $K_{ab}$, is given in Gaussian
normal coordinates by~\footnote{Note that we are using
  Misner--Thorne--Wheeler sign conventions.  In particular, see page
  552 of~\cite{MTW}.}
\begin{equation}
K_{ab} = - {1\over2} {\partial g_{ab} \over \partial n} \;.
\label{ext}
\end{equation}
It should be noted that the above curvature expressions are all
independent of the spacetime dimensionality. Nevertheless, two
transverse dimensions is somewhat special as, in this case, equation
(\ref{X1}) reduces to
\begin{equation}
{}^{(3)}R_{abcd} = 
\half \; R_\parallel \; (g_{ac} \;g_{bd} - g_{ad} \;g_{bc} ) 
 - (K_{ac} \; K_{bd} - K_{ad} \; K_{bc} ) \;.
\end{equation}
Here, we are using $R_\parallel$ to denote the Ricci scalar of the
two-dimensional surfaces of constant $n$ and $t$; that is, the Ricci
scalar of the 2-surfaces parallel to the arbitrarily chosen 2-surface.
Strictly speaking, these results have validity only on the specified
$2$-surface and in some limited region surrounding the $2$-surface;
holding only as long as the Gaussian normal coordinate system does not
break down.  (Such a breakdown tends to occur because the normal
geodesics typically intersect after a certain distance.) This is,
however, not a significant restriction on the subsequent analysis.

Let us next decompose the spatial $3$-metric in terms of the specified
$2$-metric, the set of vectors ${e^{i}{}_a}$ tangent to the
$2$-surface [the zwei-bein], and the vector $n^i$ normal to the
$2$-surface. That is,
\begin{equation}
{}^{(2+1)}g^{ij} = e^{i}{}_a \; e^{j}{}_b \; g^{ab} + n^i \; n^j \;.
\end{equation}
Also taking note of the useful identity
\begin{equation}
\tr\left({\partial K\over \partial n}\right)
= 
{\partial\tr(K)\over\partial n} -2\tr(K^2)\;,
\end{equation}
where the trace $\tr$ is performed using the $2$-metric $g_{ab}$ and its
inverse $g^{ab}$, we are able to effect the following contractions:
\begin{eqnarray}
{}^{(3)}R_{ab} &=& 
\half\; R_\parallel \; g_{ab} 
+  {\partial K_{ab} \over \partial n}  
+ 2 (K^2)_{ab} -(\tr K)\; K_{ab}\;,
\\
{}^{(3)}R_{\hat na} &=&   K_{:a}-K_{ab}{}^{:b} \;,
\\
{}^{(3)}R_{\hat n\hat n} &=& 
{\partial \tr(K) \over \partial n} - \tr(K^2)\;,
\\
{}^{(3)}R 
&=&  
R_\parallel +  2 {\partial \tr(K) \over \partial n}  - \tr(K^2) - (\tr K)^2\;.
\end{eqnarray}
We can  now evaluate the various components of the three-space Einstein
tensor ({\it cf},~\cite[page 552]{MTW}):
\begin{eqnarray}
{}^{(3)}G_{ab} &=&
{\partial K_{ab} \over \partial n}  + 2 (K^2)_{ab}  - (\tr K) K_{ab}
\nonumber\\
&& - g_{ab} \left\{ {\partial \tr(K) \over \partial n} 
- {1\over2}  \tr(K^2) - \half \; (\tr K)^2 \right\}\;,
\\
{}^{(3)}G_{\hat na} &=&   
K_{:a}-K_{ab}{}^{:b}\;,
\\
{}^{(3)} G_{\hat n\hat n} &=& 
- \half\; R_\parallel  - {1\over2} \tr(K^2) + \half (\tr K)^2 \;.
\end{eqnarray}

By way of this decomposition, we can further elaborate on the form of
the spacetime (3+1) Einstein tensor. For example,
\begin{eqnarray}
{}^{(3+1)}G_{ab} &=& 
 -  {N_{|ab}\over N} 
+ g_{ab} \left[ {{}^{(3)}\Delta N\over N} \right]
\nonumber\\
&&
+{\partial K_{ab} \over \partial n}  + 2 (K^2)_{ab} - \tr(K)\;K_{ab}
\nonumber\\
&&
+ g_{ab} \left\{- {\partial \tr(K) \over \partial n} +
{1\over2} \; \tr(K^2) +\half(\tr K)^2 \right\} \;.
\end{eqnarray}
However,  utilizing the definition of the extrinsic curvature and  the
Gauss--Weingarten equations,~\footnote{See, for instance, equations
  (21.57) and (21.63) of~\cite{MTW}. Further note that $\partial_n N=
  N_{|n}$ and these forms can be used interchangeably.}
\begin{eqnarray}
N_{|ab} &=& N_{:ab} - K_{ab} \; N_{|n} \;,
\\
N_{|na} &=&  \partial_n N_{:a} + K_{a}{}^{b} \; N_{:b} \;,
\end{eqnarray}
we then have
\begin{eqnarray}
{{}^{(3)}\Delta N} &=& 
g^{kl} \; N_{|kl} 
\\
&=& 
g^{ab} \; N_{:ab} -  (g^{ab}K_{ab}) \; N_{|n} + N_{|nn}
\\
&=& {{}^{(2)}\Delta N} - \tr(K)\; N_{|n} + N_{|nn}\;.
\end{eqnarray}
Putting everything together, we can finally write
\begin{eqnarray}
{}^{(3+1)}G_{ab} 
&=& 
 -  {N_{:ab} \over N} +  K_{ab} \; {N_{|n} \over N}
\nonumber\\
&&
+ g_{ab} \left[ {{}^{(2)}\Delta N \over N} 
+ {N_{|nn}\over N} - \tr(K)\; {N_{|n} \over N}
\right]
\nonumber\\
&&
+ {\partial K_{ab} \over \partial n}  + 2 (K^2)_{ab} - 
g_{ab} {\partial \tr(K) \over \partial n} +
{1\over2} g_{ab} \; \tr(K^2)
\nonumber\\
&&
-\tr(K)\;K_{ab} + \half(\tr K)^2 \; g_{ab} \;,
\\
{}^{(3+1)}G_{\hat na} 
&=&   
K_{:a}-  K_{ab}{}^{:b}  
-  K_a{}^b \;{N_{:b}\over N} -{ \partial_n N_{:a}\over N} \;,
\\
{}^{(3+1)}G_{\hat t a} &=&   0 \;,
\\
{}^{(3+1)} G_{\hat n\hat n} &=&  
- {1\over2} \; R_\parallel  - {1\over2} \tr(K^2) + \half(\tr K)^2 +  
{{}^{(2)}\Delta N\over N} - \tr(K)\; {N_{|n}\over N} \;,
\\
{}^{(3+1)}G_{\hat t \hat n} &=&   0 \;,
\\
{}^{(3+1)}G_{\hat t \hat t} &=&  
\half\; R_\parallel +  
{\partial \tr(K) \over \partial n}  - 
{1\over2} \tr(K^2) - \half(\tr K)^2 \;.
\end{eqnarray}
In this way, we have (for any \emph{arbitrary} static spacetime)
completely specified the (3+1)-dimensional Einstein tensor in terms of
the Ricci scalar of the \emph{arbitrarily} chosen $2$-surface, the
extrinsic geometry of this $2$-surface in the $3$-geometry of
``space'', and the ``lapse function'' $N$ (and its gradients) at the
$2$-surface.

It is sometimes useful to simplify ${}^{(3+1)}G_{ab}$ by virtue of the
fact that
\begin{eqnarray}
{N_{|nn}\over N} = - \half \; R_\perp \;,
\end{eqnarray}
where $R_\perp$ is the Ricci scalar with respect to $g_\perp$ --- the
metric in the $t$--$n$-plane perpendicular to the chosen 2-surface.

For subsequent considerations,  the following combination
is of particular interest:
\begin{eqnarray}
{}^{(3+1)}G_{\hat t \hat t} + {}^{(3+1)}G_{\hat n \hat n} 
&=& 
 {{}^{(2)}\Delta N\over N} +{\partial \tr(K) \over \partial n}  
-(\tr K) {\partial_n N\over N}
-\tr(K^2) 
\\
&=& {{}^{(2)}\Delta N\over N} 
+ N \partial_n \left[ N^{-1} \tr(K) \right] - \tr(K^2)\;.
\end{eqnarray}
Observe that the Ricci scalar $R_\parallel$ has dropped out, and so
this particular combination depends only on the extrinsic curvature
and the lapse.  Moreover, we will ultimately demonstrate that, at a
black hole horizon [static Killing horizon], this combination limits
to zero!

\section{The horizon limit}

We will now specifically consider a \emph{black hole} spacetime which,
apart from being static, is allowed to be completely general. (In
particular, we are {\em not} assuming spherical symmetry nor
asymptotic flatness, as is often the case in studies of this nature.)
The existence of a black hole horizon tells us there must be an
equipotential surface with $N=0$ ({\it i.e.}, a surface of infinite
redshift). So let us choose our arbitrary 2-surface as the $N=0$ surface, and
set up a Gaussian coordinate system $(t,n,x,y)$ with $n$ now denoting
the normal distance to the horizon.  The (3+1)-metric then takes the
particularly convenient form
\begin{equation}
g_{\mu\nu} = \left[ \begin{array}{ccc}
- N^2&0&0\\
0& 1&0\\
0& 0 & g_{ab} \end{array} \right] \;.
\end{equation}

Next, let us introduce an appropriate notion of ``local gravity''.
For this purpose, we will first define
\begin{equation}
\kappa \equiv \partial_n N 
\label{grav}
\end{equation}
and then 
\begin{equation}
\kappa_H \equiv  \lim_{n\to0} \kappa \;.
\label{gravity}
\end{equation}
One can readily verify that $\kappa_H$ complies with the standard
version of the surface gravity as defined by, for instance,
Wald~\cite{Wald}. In fact, it can also be shown that, away from the
horizon, $\kappa/N$ is simply the normal component of the
4-acceleration of an observer at fixed $n,x,y$ . (See the appendix for
further discussion.)  This enables us (in the first instance) to write
a near-horizon Taylor expansion for the lapse,
\begin{equation}
N(n,x,y) = \kappa_H(x,y)\; n + o(n^2) \;,
\end{equation}
though we will soon refine the form of this expansion considerably.

To proceed, it is useful to consider the following curvature invariant:
\begin{equation}
 {}^{(3+1)}R_{\mu\nu\alpha\beta}\;\;  {}^{(3+1)}R^{\mu\nu\alpha\beta} = 
 {}^{(3)} R_{ijkl} \;\; {}^{(3)} R^{ijkl} + 4\; {N_{|ij}\; N^{|ij}\over N^2}\;.
\end{equation}
Since we want the horizon to be regular  and \emph{not} possess a
curvature singularity, this quantity must remain finite in the horizon
limit.  Furthermore, since this is a sum of squares (relative to the
positive-definite 3-metric $g_{ij}$), it follows that the 3-geometry
must remain regular,
\begin{equation}
\lim_{n\to0} {}^{(3)} R_{ijkl} \;\; {}^{(3)} R^{ijkl}  = \hbox{finite}\;,
\end{equation}
and additionally 
\begin{equation}
\lim_{n\to0} {N_{|ij}\over N} = \hbox{finite}\;.
\end{equation}
Therefore, since the denominator is $o(n)$ for non-extremal 
horizons,~\footnote{For the time being, we will assume a non-extremal horizon
or, equivalently, that $\kappa_H$ is non-vanishing. The extremal
($\kappa_H=0$) case will be addressed in the following section.}
\begin{equation}
{N_{|ij}} = o(n) \;.
\end{equation}
Now decomposing these 3-derivatives by way of the Gauss-Weingarten
equations, we have
\begin{eqnarray}
N_{|nn}&=&o(n)\;,
\\
N_{|ab} &=& N_{:ab} - K_{ab} \; N_{|n} =o(n)\;,
\\
N_{|na} &=&  \partial_n N_{:a} + K_{a}{}^{b} \; N_{:b} = o(n)\;.
\end{eqnarray}

The first of these equations implies that we can refine the expansion
for the lapse as
\begin{equation}
N(n,x,y) = \kappa_H(x,y)\; n + o(n^3) \;.
\end{equation}
Meanwhile, the second equation indicates 
\begin{equation}
K_{ab} = o(n) \;,
\end{equation}
meaning that the extrinsic curvature limits to zero on the horizon.
Finally, the third equation implies the following:
\begin{equation}
\{\kappa_H(x,y)\}_{:a}=0 \;,
\end{equation}
which is, in fact,  the zeroth law of black hole mechanics --- the surface
gravity is a constant over the Killing horizon.  Notably, this has
been accomplished on purely geometrical grounds, without invoking any
additional constraints such as the \emph{dominant energy condition}
(DEC).  This finding is compatible with the analysis of (for instance)
Wald, where a careful inspection of pages 333-334 of~\cite{Wald}
reveals that the DEC is not actually necessary for proving the
``zeroth law'' for a static Killing horizon (although this point is
not explicitly made in the text). A further discussion  that dispenses
with the energy conditions, in more general situations than considered
in the present article, can be found in~\cite{Racz}.

The previous deductions enable us to write
\begin{equation}
N(n,x,y) = \kappa_H\;n + {\kappa_2(x,y)\over 3!} \; n^3 + o(n^4)
\end{equation}
and
\begin{equation}
g_{ab}(n,x,y) = [g_H]_{ab}(x,y) + {[g_2]_{ab}(x,y)\over2!} \; n^2 + o(n^3)
\;.
\end{equation}
Indeed, the above forms are necessary and sufficient conditions for
all polynomial scalar invariants in the Riemann tensor to be finite at
the horizon.

It is now straightforward to obtain the horizon limit for the
(3+1)-Einstein tensor:
\begin{eqnarray}
{}^{(3+1)}G_{ab}|_H 
&=& 
-\left\{ [g_2]_{ab} - [g_H]_{ab} \; \tr[g_2] \right\}
+
[g_H]_{ab} \; {\kappa_2\over\kappa_H} \;, 
\label{x64} \\
{}^{(3+1)}G_{\hat na}|_H 
&=&  0 \;,
\\
{}^{(3+1)}G_{\hat t a}|_H  &=&   0 \;,
\\
{}^{(3+1)} G_{\hat n\hat n}|_H  &=&  
- {1\over2} \; R_\parallel  + \half\; \tr[g_2] \;,
\\
{}^{(3+1)}G_{\hat t \hat n}|_H  &=&   0 \;,
\\
{}^{(3+1)}G_{\hat t \hat t}|_H  &=&  
\half\; R_\parallel -\half\; \tr[g_2]  \;.
\end{eqnarray}
We have also verified these expressions by symbolic computation.
Using the Taylor-series expansions (for the lapse and 2-metric) and
then {\em Maple} to symbolically calculate the Einstein tensor, we
have evaluated the $n\to0$ limit and found that it reproduces the
above analytic result. Similarly, we have used a {\em Maple}
calculation to verify that the horizon possesses a curvature
singularity if these Taylor series are not obeyed.~\footnote{To make
  the {\em Maple} calculation tractable, it is useful to invoke the
  remaining on-horizon coordinate freedom to write $[g_H]_{ab} =
  \exp[2\theta(x,y)] \; \delta_{ab}$.}

Of particular importance, we find (at the horizon) that
\begin{equation}
{}^{(3+1)}G_{\hat t \hat t}|_H  + {}^{(3+1)}G_{\hat n \hat n}|_H  = 0 \;,
\end{equation}
as previously advertised. Indeed, by inspection, $G_\perp \propto
g_\perp$. For the ``parallel'' [in-horizon] components, there is,
however, no generically simple relation of this type. 

Some progress may be made by first noting that, at the horizon, the
transverse Ricci scalar satisfies
\begin{equation}
 R_\perp = - 2 {N_{|nn}\over N} \to -2 {\kappa_2\over \kappa_H} \;. 
\label{x71}
\end{equation}
Next, let us split the tensor ${}^{(3+1)}G_{ab}|_H$ into a trace and
trace-free part.  Finally, using capitalized Latin indices to denote
the $t$ and $n$ directions perpendicular to the horizon, we define
$\eta_{AB} = [\hbox{diag}(-1,1)]_{AB}$. Given these considerations,
our result can then be expressed in the following compact form:
\footnote{Moreover, this version makes it particularly simple to
  analyze the situation in the special case of spherical symmetry.}
\begin{eqnarray}
{}^{(3+1)}G_{ab}|_H 
&=& 
-\half \left\{R_\perp - \tr[g_2]\right\} [g_H]_{ab}
           -\left\{ [g_2]_{ab} - \half [g_H]_{ab} \; \tr[g_2] \right\}, 
\label{blah0}
\\
{}^{(3+1)}G_{\hat Aa}|_H 
&=&  0 \;,
\label{blah1}
\\
{}^{(3+1)}G_{\hat A \hat B}|_H  &=&  -\half
\left\{ R_\parallel -\tr[g_2]\right\} \eta_{\hat A\hat B}  \;.
\label{blah2}
\end{eqnarray}
Let us again emphasize that this ``boundary condition'' holds true at
{\em any} static Killing horizon.  Moreover, given that our analysis
is purely geometrical in nature, it has quite strong repercussions on
what type of matter/energy can exist near a black hole horizon.  The
implication is that, once Einstein's equations have been imposed [{\it
  i.e.}, ${}^{(3+1)}G_{\mu\nu}=8\pi G_N \; T_{\mu\nu}\;,$ where
$T_{\mu\nu}$ is the stress-energy tensor and $G_N$ is Newton's
constant], whatever quantum fluctuations might be present are forced
to satisfy rather tight constraints. Which is to say, the stress
tensor at the horizon must take on the following block-diagonal form:
\begin{equation}
T_{\hat \mu\hat \nu}|_H  = \left[ \begin{array}{cc|c}
\rho_H &0 &0\\
0& -\rho_H & 0\\
\hline
0&  0 & T_{\hat a\hat b} \vphantom{\bigg{|}} \end{array} 
\right] \;,
\label{stress}
\end{equation}
where $\rho_H$ is the energy density at the horizon and, by virtue of
symmetry, $p_H=-\rho_H$ is the transverse component of the pressure.

That a black hole horizon enforces boundary conditions on the
curvature (and, consequently, the stress-energy) can be viewed as a
natural extension of the ``membrane paradigm'' of black hole
mechanics~\cite{membrane}.  Indeed, one can simplify many calculations
in classical black hole physics by treating the horizon as a boundary
and [typically, in the test-field limit] placing specific boundary
conditions, on say, the electric and magnetic fields~\cite{membrane}.
In this article, we have extended the notion of horizon boundary
conditions --- going beyond the test field limit --- to an arbitrary
static Killing horizon with an arbitrary matter distribution near the
horizon. More to the point, we have demonstrated that the absence of
curvature singularities at the horizon enforces a very specific
boundary condition on the curvature.

\section{Extremal horizons}

At an extremal horizon --- for which $\kappa_H=0$ --- we require a
different analysis. Let us start by assuming that the surface gravity
has an order $m$ degeneracy:
\begin{equation}
N(n,x,y) = {\kappa_m(x,y)\over (m+1)!}\; n^{m+1} + o(n^{m+2}) \;,
\end{equation}
so that
\begin{equation}
\kappa(n,x,y) = {\kappa_m(x,y)\over m!}\; n^{m} + o(n^{m+1}) \;.
\end{equation}
Then repeating the arguments that banned curvature singularities from
a non-extremal horizon, we find that the finiteness of ${}^{(3+1)}G_{na}$
requires that
\begin{equation}
N(n,x,y) = {\kappa_m\over (m+1)!}\; n^{m+1} 
+ {\kappa_{m+2}(x,y)\over (m+3)!}\; n^{m+3} + o(n^{m+4})
\end{equation}
and 
\begin{equation}
\kappa(n,x,y) = {\kappa_m\over m!}\; n^{m} 
+  {\kappa_{m+2}(x,y)\over (m+2)!}\; n^{m+2} + o(n^{m+3}) \;.
\end{equation}
Unfortunately, there is now a problem with ${}^{(3+1)}G_{ab}$:
\begin{equation}
{}^{(3+1)}G_{ab} = g_{ab} {N_{|nn}\over N} + o(1) = { m(m+1)\over n^2} g_{ab} +  o(1),
\end{equation}
indicating that the horizon has a curvature singularity unless
$m(m+1)=0$. Now $m=0$ corresponds to the non-extremal horizon
considered earlier, while $m=-1$ corresponds to the lapse being finite
and no horizon forming.

Thus if an extremal horizon is located at any finite value of the
normal coordinate $n$ (so that it makes sense to shift the location to
$n=0$), we must conclude that the extremal horizon possesses a
curvature singularity. Conversely, any extremal horizon that is not
simultaneously a curvature singularity must be located at infinite
proper distance $n=-\infty$. It is gratifying to see this well-known
result regarding extremal horizons~\cite{extremal} arising naturally
in this new context.

\section{Conformal deformations}

Let us now investigate what happens to the on-horizon structure of the
Einstein-tensor under a conformal deformation of the spacetime metric,
\begin{equation}
g_{\mu\nu} \to \tilde g_{\mu\nu}= \exp[2\,\omega(n,x,y)] \; g_{\mu\nu} \;,
\end{equation}
with
\begin{equation}
\omega(n,x,y) = \omega_H(x,y) +\omega_1(x,y) \; n 
+ \half\omega_2(x,y) \; n^2 + O(n^3) \;.
\end{equation}
For any such deformation, Jacobson and Kang~\cite{JK} have
demonstrated that the notion of a horizon and the value of the surface
gravity, $\kappa_H$, are invariants. Indeed, if $\omega$ is time
independent (as assumed here), then the deformed geometry $\tilde
g_{\mu\nu}$ has the same timelike Killing vector as the original
geometry $g_{\mu\nu}$. Consequently, the \emph{form} of the Einstein
tensor (and, in fact, all of the curvature tensors) will be unaltered.
The specific \emph{value}, however, will generally change.  (That is,
the form of the on-horizon curvature is a conformal invariant, while
the value of the on-horizon curvature is conformally covariant.)

Actually, a direct application of our prior formalism is a little
tricky since we would first need to construct [to suitable accuracy]
the Gaussian normal coordinate patch $(t,\tilde n, \tilde x, \tilde
y)$ appropriate to the metric $\tilde g_{\mu\nu}$, and such a
construction is a tedious exercise that is quite prone to error.
Instead, we will start with the well-known transformation law for the
Einstein tensor under an arbitrary conformal transformation
\cite{Wald},
\begin{equation}
{}^{(3+1)}\tilde G_{\mu\nu} = {}^{(3+1)}G_{\mu\nu} 
-2 \nabla_\mu\nabla_\nu \omega + 2\nabla_\mu \omega\; \nabla_\nu \omega
+g_{\mu\nu} \left\{ 2\; {}^{(3+1)} \Delta \omega 
+ (\nabla\omega)^2 \right\} \;,
\end{equation}
and confront it with the symmetries (as previously established) of the
Einstein tensor at any static Killing horizon.

First of all, the preservation of these symmetries [in particular,
equations (\ref{blah1}) and (\ref{blah2})] immediately implies that
\begin{equation}
\lim_{n\to0} \left\{ \omega_{|nn} - \omega_{|n} \; \omega_{|n} \right\} = 0
\end{equation}
and
\begin{equation}
\lim_{n\to0} \left\{ \omega_{|na} - \omega_{|n} \; \omega_{|a} \right\} = 0
\;.
\end{equation}
One effect of these constraints is to enforce $\omega_2=(\omega_1)^2$. But
if $\omega_1\neq0$, then the quantity
\begin{equation}
{}^{(3+1)}\Delta \omega = 
{1\over N \sqrt{\det[g_{ab}]}} \; 
\partial_\mu \left( N \sqrt{\det[g_{ab}]} \; 
g^{\mu\nu} \; \partial_\nu \omega\right) 
\end{equation}
simplifies to
\begin{equation}
{}^{(3+1)}\Delta \omega = {}^{(2)}\Delta \omega + 
{1\over \kappa_H \; n} \partial_n 
\left(  \kappa_H \; n [\omega_1 + \omega_2\; n] \right) + o(n) 
= {\omega_1\over n} + o(1) \to \infty \; ;
\end{equation}
thus leading to an undesirable curvature singularity at the horizon.
Therefore, we must have $\omega_1=0$ and, consequently, $\omega_2=0$.
[In contrast, $\omega_H(x,y)$ remains an arbitrary unconstrained
function.] Hence,
\begin{equation}
\omega(n,x,y) = \omega_H(x,y) + O(n^3) \;,
\end{equation}
and the on-horizon limit now  yields
\begin{equation}
{}^{(3+1)}\Delta \omega = {}^{(2)}\Delta \omega + 
{1\over \kappa_H \; n} \partial_n \left(  o(n^3)  \right) + o(n) 
\to {}^{(2)}\Delta \omega_H \;,
\end{equation}
which is finite. Similarly,
\begin{equation}
\lim_{n\to0} \nabla_a\nabla_b 
\omega \equiv \lim_{n\to0} \omega_{|ab} =  \{\omega_H\}_{:ab}\;,
\end{equation}
and
\begin{equation}
\lim_{n\to0} \nabla_a \omega\; \nabla_b \omega 
= \{\omega_H\}_{:a} \; \{\omega_H\}_{:b}\;,
\end{equation}
so that,  for the on-horizon Einstein tensor, we have
\begin{eqnarray}
{}^{(3+1)}\tilde G_{ab}|_H 
&=& {}^{(3+1)}G_{ab}|_H - 2  \{\omega_H\}_{:ab} 
+ 2  \{\omega_H\}_{:a} \; \{\omega_H\}_{:b}
\nonumber\\
&&\qquad
+ g_{ab} \left\{ 2 \; {}^{(2)}\Delta 
\omega_H + g^{cd} \; \{\omega_H\}_{:c} \; \{\omega_H\}_{:d} \right\}\;,
\label{x92} \\
{}^{(3+1)}\tilde G_{\hat Aa}|_H 
&=&  0 \;,
\\
{}^{(3+1)}\tilde G_{\hat A \hat B}|_H  &=&  \exp(-2\omega_H) \Bigg[
{}^{(3+1)} G_{\hat A \hat B}|_H  
\nonumber\\
&&\qquad 
+\eta_{\hat A\hat B} \left\{ 2\;  {}^{(2)}\Delta \omega_H + g^{cd} \; 
\{\omega_H\}_{:c} \; \{\omega_H\}_{:d} \right\} 
\Bigg]  \;.
\label{x94} 
\end{eqnarray}
Let us be clear about what has transpired here.  The form of the
conformal deformation near the horizon is constrained by the need to
avoid curvature singularities at the horizon. Once this has been done,
the \emph{change} in the Einstein tensor at the horizon depends only
on the conformal deformation at the horizon itself.

The results of this section have, once again, been checked by symbolic
computation.  To elaborate, we have used the Taylor series for
$\omega(n,x,y)$ to set up a {\em Maple} computation of the Einstein
tensor in the vicinity of the horizon and then symbolically taken the
$n\to0$ limit. An independent computation that employs the $(t,\tilde
n, \tilde x, \tilde y)$ Gaussian coordinate patch appropriate to the
metric $\tilde g_{\mu\nu}$ is too tedious to be worth presenting in
any detail. It is, however, relatively simple to work backwards from
\begin{equation}
R_\parallel(\tilde g_H) =  \exp(-2\omega_H) 
\left\{ R_\parallel(g_H) - 2\;  {}^{(2)}\Delta 
\omega_H \right\}
\end{equation}
to find that [{\it cf}, equations (\ref{blah2}) and (\ref{x94})]
\begin{equation}
\tilde{\tr}[\tilde g_2] =  \exp(-2\omega_H) 
\left\{ \tr[ g_2] +  2 \; {}^{(2)}\Delta \omega_H + 2 \; g^{cd} \; 
\{\omega_H\}_{:c} \; \{\omega_H\}_{:d} \right\}\;.
\end{equation}
Furthermore, 
from the trace-free part of ${}^{(3+1)}\tilde G_{ab}|_H$, we can extract
 [using equations (\ref{blah0}) and (\ref{x92})]
\begin{eqnarray}
&&\left\{ 
[\tilde g_2]_{ab} - \half [\tilde g_H]_{ab} \; \tilde{\tr}[\tilde g_2] 
\right\}
=
\left\{ [g_2]_{ab} - \half [g_H]_{ab} \; \tr[g_2] \right\}
\\
&&
+ 2  \left[   \{\omega_H\}_{:ab} 
-  \{\omega_H\}_{:a} \; \{\omega_H\}_{:b} - 
\half g_{ab} \left(
{}^{(2)}\Delta \omega_H - g^{cd} \; 
\{\omega_H\}_{:c} \; \{\omega_H\}_{:d} 
\right) \right] \;.
\nonumber
\end{eqnarray}
As a consequence, it can
also be shown that [{\it cf}, equation(\ref{x64})]
\begin{equation}
[\tilde \kappa_2] =  \exp(-2\omega_H) \left\{ [\kappa_2] + 
\kappa_H \; g^{cd} \; 
\{\omega_H\}_{:c} \; \{\omega_H\}_{:d} 
\right\}.
\end{equation}
This last equation can equivalently be written [via equation (\ref{x71})] 
\begin{equation}
R_\perp(\tilde g_H) =  \exp(-2\omega_H) 
\left\{ R_\perp(g_H) - 2\;  
g^{cd} \; 
\{\omega_H\}_{:c} \; \{\omega_H\}_{:d} 
 \right\}.
\end{equation}
The (3+1)-dimensional conformal transformation is actually changing
the location of the transverse surfaces at the order $o(n^2)$, and so
this transformation law for $R_\perp$ is what one might have naively
expected. In any event, we have confronted this calculation with an
explicit coordinate computation that employs Gaussian coordinates,
again verifying these formulae to be correct.

\section{Discussion}

To summarize, we have demonstrated that the Einstein tensor must take
on a particularly simple form [see equations
(\ref{blah0})--(\ref{blah2})] at any non-extremal static Killing
horizon.  In particular, we have shown that the on-horizon Einstein
tensor block-diagonalizes and, moreover, that ${}^{(3+1)}G_{\hat t
  \hat t}|_H + {}^{(3+1)}G_{\hat n \hat n}|_H = 0\; $; where $\hat t$
and $\hat n$ are the (normalized) spacetime coordinates in the
directions perpendicular to the horizon (timelike and spacelike
respectively).  Although this symmetry had already been established
for static spherically symmetric horizons~\cite{DBH1,PAD}, this is ---
to the best of our knowledge --- the first time that it has been
rigorously demonstrated to hold for generic static geometries.

A direct implication of our analysis is that, by way of Einstein's
equation, the matter/energy near a static Killing horizon
(\emph{including any quantum fluctuations}) must be highly
constrained; {\it cf}, equation (\ref{stress}). Furthermore, we can
now make the following observation: Given that the ``interesting''
near-horizon physics can be anticipated to take place in the
$n$--$t$ plane~\cite{CAR3} and $T_\perp \propto g_\perp$ at the
horizon, the near-horizon stress tensor is effectively that of a
collection of world-sheet conformal field theories.  More precisely, a
collection of two-dimensional conformal theories, each of which is
defined at a point on the horizon and constrained to act in the
$n$--$t$ plane.  [And with interactions between these conformal field
theories possibly being responsible for the in-horizon portion of the
(3+1)-stress-energy tensor.]  One can now see how two-dimensional
conformal theories could play such a prominent role in calculations of
black hole entropy, as has been central to the program of
Carlip~\cite{CAR1,CAR2,CAR4}, as well as Solodukhin~\cite{SOL}.
Moreover, we would suggest that it is these geometrical constraints on
the stress tensor that underlie Carlip's notion of black hole entropy
being controlled by an asymptotic conformal symmetry near the horizon.

With regard to our last (somewhat speculative) remark, it may be
significant that {\em any} static conformal transformation will not
alter the general form of the Einstein or stress tensor, and the
conformal deformation of these tensors will be highly constrained
({\it cf}, Section 5).  Coupled with the knowledge that such a
transformation also preserves the causal structure of the solution
\cite{JK}, it becomes evident that our geometric symmetry is truly
conformal in its nature.

An interesting open question is what (if any) symmetries would persist
when the Killing horizon is no longer static.  Although a technically
difficult problem, we do anticipate that stationary Killing horizons
would exhibit similar symmetries to those found for the static case;
and work along this direction is currently underway.  Meanwhile, a
truly time-dependent geometry raises serious issues that go beyond
mere technical difficulties.  On the other hand, one might expect
that, if a horizon is evolving slowly enough, it should have a viable
interpretation as being quasi-static (or, at least, quasi-stationary).
Under such circumstances, our current formalism could still be applied
to time-dependent scenarios with some degree of accuracy.

\section*{Acknowledgements}

Research supported by the Marsden Fund administered by the Royal
Society of New Zealand, and by the University Research Fund of
Victoria University.

\section*{Appendix: Defining the surface gravity}

Here, we will verify that our version of the surface gravity [{\it
  cf}, equations (\ref{grav}) and (\ref{gravity})] is indeed
compatible with the standard one (see for instance Wald~\cite{Wald}).
Let us start by noting that, if $\xi^{\mu}$ is the Killing vector for
our static spacetime, then
\begin{equation}
\xi^{\mu}\xi_{\mu}= -N^2 \;.
\end{equation}
Hence, the $4$-velocity of the static fiducial observers [FIDOS] can
be defined by
\begin{equation}
V^{\mu}= {\xi^{\mu}\over N} \;.
\end{equation}
Consequently,  the $4$-acceleration of the FIDOS is
given by
\begin{equation}
A^{\mu} = (V^{\nu}\nabla_{\nu}) V^{\mu} = {1\over N} (\xi^{\nu}\nabla_{\nu}) 
\left[{\xi^{\mu}\over N}\right] 
= {1\over N^2}  (\xi^{\nu}\nabla_{\nu}) \xi^{\mu} - {\xi^{\mu}\over N^3} 
(\xi^{\nu} 
\nabla_{\nu}) N 
= {(\xi^{\nu} \nabla_{\nu}) \xi^{\mu}\over N^2},
\end{equation}
where only at the last step has the fact that $\xi^{\mu}$ is a Killing
vector been invoked. Next, let us consider that
\begin{equation}
A_{\mu} = {(\xi^{\nu} \nabla_{\nu}) \xi_{\mu}\over N^2} 
= - {(\xi^{\nu}\nabla_{\mu} ) \xi_{\nu}\over N^2} =
-\half\;{\nabla_{\mu} (\xi^{\nu}\xi_{\nu}) \over N^2} \;.
\end{equation}
Therefore,
\begin{equation}
A_{\mu} = +\half \; {\nabla_{\mu} (N^2)\over N^2} = {\nabla_{\mu} N\over N} \;,
\end{equation}
or, to put it another way,
\begin{equation}
||A||={||\nabla N ||\over N}\;.
\end{equation}

Let us now recall equation (\ref{grav}) for our definition of the
``local gravity''.  In view of this definition, as well as the acceleration
being normal to surfaces of constant $N$, it follows that
\begin{equation}
\kappa = \partial_n N = N \; ||A|| \;,
\end{equation}
and so
\begin{equation}
\kappa_H 
= \lim_{z\to H} \partial_n N 
= \lim_{z\to H} \left\{N \; ||A|| \right\}
\;,
\end{equation}
in agreement with equation (12.5.18) of~\cite{Wald}.  Moreover, since
equation (12.5.18) was derived directly from equation (12.5.2)
of~\cite{Wald} (the latter being Wald's starting-point definition of
the surface gravity), it is clear that our definition of $\kappa_H$
complies with the standard version.




\begin{thebibliography}{99}

\bibitem{BEK} J.D. Bekenstein, ``Black holes and the second law'',
Lett. Nuovo Cim. {\bf 4}, 737 (1972); \\
``Black holes and entropy'', Phys. Rev. D {\bf 7}, 2333 (1973); \\
``Generalized second law of thermodynamics in black hole physics'',
Phys. Rev. D {\bf 9}, 3292 (1974).

\bibitem{HAW} S.W. Hawking,
``Black Hole Explosions'',
Nature {\bf 248} (1974) 30;\\
``Particle creation by black holes'',
Commun. Math. Phys. {\bf 43}, 199 (1975).

\bibitem{DAM} See, for a recent review and references, \\
              T. Damour, ``The entropy of black holes: a primer'',
              arXiv:hep-th/0401160 (2004).

\bibitem{SMO} See,  for a relevant discussion,\\
               L. Smolin, ``How far are we from the theory of quantum 
               gravity'', arXiv:hep-th/0303185 (2003).

\bibitem{STR-V} A. Strominger and C. Vafa, ``The Microscopic origin of
black hole entropy'', Phys. Lett. B {\bf 379}, 99 (1996)
[arXiv:hep-th/9601029].

\bibitem{ASH}
A. Ashtekar, J. Baez, A. Corichi and K. Krasnov,
``Quantum Geometry and Black Hole Entropy'',
Phys. Rev. Lett. {\bf 80}, 904 (1998)
[arXiv:gr-qc/9710007].


\bibitem{CAR3} S. Carlip, ``Black hole entropy from
horizon conformal field theory'', Nucl. Phys. Proc. Suppl. {\bf 88},
10 (2000) [arXiv:gr-qc/9912118].


\bibitem{STR} A. Strominger, ``Black hole entropy from near-horizon
microstates'', JHEP {\bf 9802}, 009 (1998) [arXiv:hep-th/9712251].

\bibitem{BTZ} M. Banados, J. Teitelboim and J. Zanelli, ``The black
hole in three dimensional space time'', Phys. Rev. Lett. {\bf 69},
1849 (1992) [arXiv:hep-th/9204099].


\bibitem{CARDY} J.L. Cardy, ``Operator content of two-dimensional
conformally invariant theories'',  Nucl. Phys. B {\bf 270}, 186 (1986).


\bibitem{BH} J.D. Brown and M. Henneaux,
``Central charges in the canonical realization of asymptotic
symmetries: An example from three-dimensional gravity'',  
 Commun. Math. Phys. {\bf 104},
207 (1986).

\bibitem{CAR1} S. Carlip, ``Black hole entropy from
conformal field theory in any dimension'',
Phys. Rev. Lett. {\bf 82}, 2828 (1999) 
[arXiv:hep-th/9812013].


\bibitem{CAR2} S. Carlip, ``Entropy from conformal field theory at
Killing horizons'', Class. Quant. Grav. {\bf 16}, 3327 (1999)
[arXiv:gr-qc/9906126].


\bibitem{CAR4} S. Carlip, ``Near-Horizon conformal symmetry and black
hole entropy'', Phys. Rev. Lett. {\bf 88}, 241301 (2002)
[arXiv:gr-qc/0203001].

\bibitem{SOL} S.N. Solodukhin, ``Conformal description of horizon's states'',
Phys. Lett. B {\bf 454}, 213 (1999) [arXiv:hep-th/9812056].



\bibitem{DBH1} M. Visser, ``Dirty black holes: Thermodynamics
and horizon structure'', Phys. Rev. D {\bf 46}, 2445 (1992)
[arXiv:hep-th/9203057].


\bibitem{PAD} T. Padmanabhan, ``Classical and quantum thermodynamics
of horizons in spherically symmetric spacetimes'', Class. Quant. Grav.
{\bf 19}, 5387 (2002) [arXiv:gr-qc/0204019].



\bibitem{DBH2} M. Visser, ``Dirty black holes: Entropy versus area'',
Phys. Rev. D {\bf 48}, 583 (1993)
[arXiv:hep-th/9303029].


\bibitem{DBH3} M. Visser, ``Dirty black holes: entropy as a
surface term'', Phys. Rev. D {\bf D8}, 5697 (1993)
[arXiv:hep-th/9307194].


\bibitem{Wald} 
R.M. Wald, 
{\em General Relativity}, 
(University of Chicago Press, Chicago, 1984).

  
\bibitem{Israel67}
W. Israel, ``Event horizons in static vacuum space-times'', 
Phys. Rev. {\bf 164}  1776 (1967).

\bibitem{Israel68}
W. Israel,  ``Event horizons in static electrovac spacetimes'',
Commun. Math. Phys. {\bf 8}, 254  (1968). 

\bibitem{Israel86}
W. Israel, ``The formation of black holes in nonspherical
collapse and cosmic censorship'', 
Can. J. Phys. {\bf 64}, 120  (1986). 


\bibitem{Hochberg-Visser}
D. Hochberg and M. Visser, 
``Geometric structure of the generic static traversable wormhole throat'',
Phys. Rev. D {\bf 56}, 4745 (1997)
[arXiv:gr-qc/9704082].

\bibitem{Hochberg-Visser-2}
D. Hochberg and M. Visser, ``Generic wormhole throats'',
in Proceedings of the Haifa Workshop: The Internal Structure of Black
Holes and Spacetime Singularities (Haifa, Israel, June-July 1997)
[arXiv:gr-qc/9710001].

\bibitem{MTW} 
C.W. Misner, K.S. Thorne and J.A. Wheeler, 
{\em Gravitation} 
(W.H. Freeman, San Francisco, 1973).

\bibitem{Hawking-Ellis} 
S.W. Hawking and G.F.R. Ellis,
{\em The Large Scale Structure of Space-Time},
(Cambridge University Press, Cambridge, England, 1973).

\bibitem{Racz}
I.~Racz and R.~M.~Wald,
 ``Global extensions of space-times describing asymptotic 
final states of black holes,''
arXiv:gr-qc/9507055 (1995).

\bibitem{membrane} 
K.S.~Thorne, R.H.~Price, and  D.A.~Macdonald (editors),
  {\it Black Holes: the membrane paradigm}, 
(Yale University Press, New Haven, 1986).


\bibitem{extremal}
See for instance,
\\
I.~Racz,
 ``Does the third law of black hole thermodynamics really have a serious
failure?,''
Class.\ Quant.\ Grav.\  {\bf 17} (2000) 4353
[arXiv:gr-qc/0009049].
\\
F.~Bonjour and R.~Gregory,
 ``Comment on 'Abelian Higgs hair for extremal black holes and selection  rules
for snapping strings',''
Phys.\ Rev.\ Lett.\  {\bf 81} (1998) 5034
[arXiv:hep-th/9809029].


\bibitem{JK} 
T. Jacobson and G. Kang, ``Conformal invariance of black
hole temperature'', Class. Quant. Grav. {\bf 10}, L201
(1993) [arXiv:gr-qc/9307002].



\end{thebibliography}
\end{document}